\documentclass[journal]{IEEEtran}
\usepackage[utf8]{inputenc}
\usepackage{amsmath}
\usepackage{soul}
\usepackage{array}
\usepackage{amsmath,amsthm,amssymb,epsf}
\usepackage{bm} 
\usepackage{graphicx,psfrag}
\usepackage{epstopdf}
\usepackage{verbatim}
\usepackage{multirow}
\usepackage{color}
\usepackage{setspace}
\usepackage{enumitem}
\usepackage{subfigure}
\usepackage{amssymb}
\usepackage{comment}
\usepackage{tabularx}
\usepackage{booktabs}
\usepackage{dirtytalk}
\usepackage{amsmath}
{
      \theoremstyle{plain}
      
  }
\usepackage{csquotes}

\newtheorem{definition}{Definition}

\newtheorem{lemma}{Lemma}

\newtheorem{remark}{Remark}

\usepackage{soul}
\usepackage{cite}

\usepackage{algorithmicx}
\usepackage[ruled,vlined, linesnumbered]{algorithm2e}

\makeatletter
\newcommand{\removelatexerror}{\let\@latex@error\@gobble}

\IEEEoverridecommandlockouts
\begin{document}

\title{Online Reduced-Order Data-Enabled Predictive Control}

\author{Amin~Vahidi-Moghaddam, Kaixiang~Zhang, Xunyuan Yin, Vaibhav~Srivastava, and Zhaojian Li
\thanks{This work was partially supported by the U.S. National Science Foundation Award CMMI-2320698 and the U.S. Army DEVCOM Ground Vehicle Systems Center (GVSC) under Cooperative Agreement W56HZV-19-2-0001.}
\thanks{Amin Vahidi-Moghaddam, Kaixiang~Zhang, and Zhaojian Li are with the Department of Mechanical Engineering, Michigan State University, East Lansing, MI 48824 USA (e-mail: vahidimo@msu.edu,  zhangk64@msu.edu, lizhaoj1@egr.msu.edu).}
\thanks{Xunyuan Yin is with the School of Chemistry, Chemical Engineering and Biotechnology, Nanyang Technological University, Singapore (e-mail: xunyuan.yin@ntu.edu.sg)}
\thanks{Vaibhav Srivastava is with the Department of Electrical and Computer Engineering, Michigan State University, East Lansing, MI 48824 USA (e-mail: vaibhav@egr.msu.edu)}
} 

\maketitle

\begin{abstract}
Data-enabled predictive control (DeePC) has garnered significant attention for its ability to achieve safe, data-driven optimal control without relying on explicit system models. Traditional DeePC methods use pre-collected input/output (I/O) data to construct Hankel matrices for online predictive control. However, in systems with evolving dynamics or insufficient pre-collected data, incorporating real-time data into the DeePC framework becomes crucial to enhance control performance. This paper proposes an online DeePC framework for time-varying systems (i.e., systems with evolving dynamics), enabling the algorithm to update the Hankel matrix online by adding real-time informative signals. By exploiting the minimum non-zero singular value of the Hankel matrix, the developed online DeePC selectively integrates informative data and effectively captures evolving system dynamics. Additionally, a numerical singular value decomposition technique is introduced to reduce the computational complexity for updating a reduced-order Hankel matrix. Simulation results on two cases, a linear time-varying system and the vehicle anti-rollover control, demonstrate the effectiveness of the proposed online reduced-order DeePC framework.
\end{abstract}


\IEEEpeerreviewmaketitle

\section{Introduction}
With the increasing complexity of modern control systems and the expanding availability of data, there is a growing preference for data-driven control schemes \cite{chen2024data}. Different from model-based control schemes that require precise system modeling, data-driven controllers generate control policies by using collected input/output (I/O) data \cite{wang2024data,hui2024sampled,huang2024data}. Data-driven control can be categorized into two paradigms: indirect data-driven control, which includes system identification and model-based control design \cite{vahidi2020memory}, and direct data-driven control, which bypasses the system identification process and directly designs control strategies based on I/O data \cite{fu2023data}. The indirect data-driven control provides an approximated model when a first-principle model is either unavailable or too difficult/costly to obtain. However, the direct data-driven control offers greater flexibility and potential compared to the indirect data-driven control by eliminating the need for a specific parametric model in control design \cite{de2019formulas}.

Recently, a direct data-driven control approach,  known as data-enabled predictive control (DeePC) \cite{coulson2019data}, has received extensive attention. DeePC employs Fundamental Lemma \cite{willems2005note}, which is a result from behavioral systems theory \cite{willems1997introduction}, to describe system behaviors in a non-parametric manner. Unlike existing approaches that involve identifying or learning a parametric model to represent the system dynamics, the Fundamental Lemma characterizes the entire space of the system's I/O trajectories using a Hankel matrix constructed from pre-collected I/O data. This results in a non-parametric model that can describe system behavior if a rank condition on the Hankel matrix is satisfied. Compared to machine learning approaches, the Fundamental Lemma methods are more computationally efficient, less data hungry, and more suitable for rigorous stability and robustness analysis \cite{carlet2022data,talebzadeh2024deep,boroujeni2024comprehensive}. However, the Fundamental Lemma only holds for deterministic linear time-invariant (LTI) systems \cite{coulson2022quantitative,teutsch2023online}. For other systems such as nonlinear systems, stochastic systems, and time-varying systems, the rank condition on the Hankel matrix is not sufficient to accurately determine the trajectory subspace, which may lead to poor performance in data-driven control. Several techniques employing slack variables and regularization methods have been proposed to address these limitations and enhance performance for the nonlinear systems and the stochastic systems \cite{coulson2019data,elokda2021data,berberich2021data}. Moreover, a robust Fundamental Lemma has been proposed to ensure the persistently exciting (PE) input with sufficient order for the stochastic LTI systems \cite{coulson2022quantitative}. However, these techniques are only effective in local regions captured by the pre-collected I/O data and cannot handle the new system dynamics that emerge online. Therefore, it is necessary to update the Hankel matrix online using real-time I/O data to predict system behavior accurately. 

For time-varying systems, i.e., systems with evolving dynamics, online DeePC \cite{baros2022online,berberich2022linear} is developed to continuously update the Hankel matrix using real-time data. In \cite{baros2022online}, old data is replaced with new data: the first column (the oldest data point) is discarded from the Hankel matrix, all columns are shifted back by one step, and the most recent real-time data is added as the last column. However, this method requires a PE control input in real-time, which is achieved by adding a suitable excitation, such as injecting noise into the control input during closed-loop operation. Injecting noise can deteriorate control performance and put unnecessary stress on actuators. To address this, \cite{teutsch2023online} presents a discontinuous online DeePC method that replaces the PE requirement with a rank condition on mosaic-Hankel matrix (a Hankel matrix with discontinuous I/O trajectories) proposed by \cite{markovsky2022identifiability}. This approach requires only an offline PE input trajectory, which is produced based on the robust Fundamental Lemma \cite{coulson2022quantitative}, and updates the mosaic-Hankel matrix if the rank condition is satisfied. However, as mentioned before, a rank condition (or PE condition) is not sufficient to indicate informative data for non-deterministic LTI systems.

\cite{shi2023efficient} proposes a continuous online DeePC by adding real-time I/O trajectories to the Hankel matrix, which removes the PE requirement for the real-time control input as the rank condition is always satisfied. However, continuously increasing the columns of the Hankel matrix leads to high memory and computational costs. To mitigate this, two numerical singular value decomposition (SVD) algorithms are used alternatively based on the rank of the Hankel matrix: when the rank of the Hankel matrix is less than the number of rows, the numerical SVD algorithm \cite{brand2006fast}, which is designed based on the eigen-decomposition algorithm \cite{gu1993stable}, is applied; and when the rank of the Hankel matrix is equal to the number of rows, the numerical SVD algorithm \cite{bunch1978updating} is employed. The online DeePC scheme \cite{shi2023efficient} faces three main limitations: (i) it requires adding all real-time trajectories to the Hankel matrix, leading to high computational cost and the inclusion of data without considering its informativeness; (ii) the numerical SVD algorithm \cite{bunch1978updating}, as mentioned in \cite{gu1993stable}, is not fast and can be unstable; and (iii) the dimension of the online SVD-based DeePC can be reduced to a minimum possible dimension. These limitations motivate us to propose a new online reduced-order DeePC. Our approach measures the informativeness of real-time data by observing the minimum non-zero singular value of the Hankel matrix and uses the most informative data discontinuously, which addresses challenge (i). By setting a well-tuned threshold on the minimum non-zero singular value, we are able to capture the new system dynamics that emerge online and update the Hankel matrix only with the most informative data. Moreover, to overcome challenges (ii) and (iii), we modify the numerical SVD algorithm \cite{brand2006fast}, as this algorithm only works for low-rank modifications, and develop an online reduced-order DeePC with adaptive order. These modifications result in lower computational complexity for the online DeePC and may improve the control performance for some cases.

The remainder of this work is structured as follows. Section II gives an introduction to the Fundamental Lemma, the discontinuous Fundamental Lemma, and the DeePC framework. In Section III, we present the proposed online DeePC algorithm and online reduced-order DeePC algorithm. In Section IV, we evaluate the developed control scheme via simulation studies on a linear time-varying system and the vehicle anti-rollover control problem. The paper concludes in Section V.

\textbf{Notations}. We adopt the following notations across the paper. $\mathbb{R}^n$ and $\mathbb{R}^{n\times m}$ represent the set of $n$-dimensional real vectors and the set of $n\times m$-dimensional real matrices, respectively. $x^{\top}$ and $A^{\top}$ stand for the transpose of the vector $x$ and the matrix $A$, respectively. $\lVert \cdot \rVert$ denotes the Euclidean norm of a vector or the induced $2$-norm of a matrix. $\rm{rank}(A)$ represents the rank of $A$. $\sigma_{\rm{min}} (A)$ and $\sigma_{\rm{max}} (A)$ stand for the minimum singular value and the maximum singular value of $A$, respectively. $\sigma_{\rm{r}} (A)$ represents the corresponding singular values to $\rm{rank}(A)$, which is the minimum non-zero singular value of matrix $A$. $I_n$ stands for the identity matrix with $n$-dimension.

\section{Preliminaries and Problem Statement}
\label{Sec2}
Consider the following unknown discrete-time system:
\begin{equation}
  \begin{aligned}
    \label{system}
     x(k+1) &= f(x(k),u(k)),\\
     y(k) &= h(x(k),u(k)),
  \end{aligned}
\end{equation}
where $k \in \mathbb{N}^+$ represents the time step, $x \in \mathbb{R}^n$ denotes the system state, $u \in {\mathbb{R}^m}$ indicates the control input, and $y \in \mathbb{R}^p$ stands for the system output. Moreover, $f:\mathbb{R}^n\times \mathbb{R}^m \rightarrow \mathbb{R}^n$ indicates the system dynamics with $f(0,0)=0$, and $h:\mathbb{R}^n\times \mathbb{R}^m \rightarrow \mathbb{R}^p$ denotes the output dynamics.

Assuming the system \eqref{system} is a linear time-invariant (LTI) system, i.e., $f(x(k),u(k))=A x(k) + B u(k)$ and $h(x(k),u(k)) = C x(k) + D u(k)$, the Fundamental Lemma \cite{willems2005note} provides a non-parametric representation to describe the system behavior.

\begin{definition}[Hankel Matrix]
\label{Hankel Matrix}
Given a signal $w(k)\in \mathbb{R}^q$, we denote by $w_{1:T}$ the restriction of $w(k)$ to the interval $[1, T]$, namely $w_{1:T} = [w^{\top}(1), w^{\top}(2), \cdots, w^{\top}(T)]^{\top}$. The Hankel Matrix of depth $K \leq T$ is defined as:
\begin{equation}
  \begin{aligned}
    \label{Hankel w}
    & \mathcal{H}_K(w_{1:T}):= \begin{bmatrix}
    w(1) & w(2) & \cdots & w(T-K+1)\\ 
    w(2) & w(3) & \cdots & w(T-K+2)\\
    \vdots & \vdots & \ddots & \vdots\\ 
    w(K) & w(K+1) & \cdots & w(T)
    \end{bmatrix}.
  \end{aligned}
\end{equation}
Let $L:= T-K+1$, then we have $\mathcal{H}_K(w_{1:T})\in \mathbb{R}^{qK\times L}$.
\end{definition}

\begin{definition} [Persistently Exciting]
	The sequence $w_{1:T}$ is persistently exciting (PE) of order $K$ if $\mathcal{H}_{K}(w_{1:T})$ has full row rank, i.e., $rank(\mathcal{H}_{K}(w_{1:T})) = qK$. 
\end{definition}

\begin{lemma}[Fundamental Lemma \cite{willems2005note}]\label{Fundamental Lemma}
	Consider the system \eqref{system} as a controllable LTI system with a pre-collected input/output (I/O) sequence $(u_{1:T}^{\mathrm{d}}, y_{1:T}^{\mathrm{d}})$ of length $T$. Providing a PE input sequence $u^{\mathrm{d}}_{1:T}$ of order $K+n$, any length-$K$ sequence  $(u_{1:K}, y_{1:K})$ is an I/O trajectory of the LTI system if and only if we have
	\begin{equation} \label{fundamental lemma}
	{
		\begin{bmatrix}
			u_{1:K} \\ y_{1:K}
		\end{bmatrix} = \begin{bmatrix}
			\mathcal{H}_{K}(u^{\mathrm{d}}_{1:T}) \\
			\mathcal{H}_{K}(y^{\mathrm{d}}_{1:T})
		\end{bmatrix} g,
	}
	\end{equation}
	for some real vector $g \in \mathbb{R}^L$.
\end{lemma}

\begin{remark} [Rank of Hankel Matrix]
\label{Rank of Hankel Matrix}
Considering $r:= \rm{rank}\left(\begin{bmatrix}
			\mathcal{H}_{K}(u^{\mathrm{d}}_{1:T}) \\
			\mathcal{H}_{K}(y^{\mathrm{d}}_{1:T})
		\end{bmatrix}\right)$, a persistently exciting control input sequence of order $K+n$ ensures that $mK+1 \leq r \leq mK+n$. Furthermore, $r = mK+n$ if $K \geq l$, where $l \leq n$ is the observability index. See \cite{markovsky2021behavioral} for more details.
\end{remark}

The Fundamental Lemma shows that the Hankel matrix in~\eqref{fundamental lemma} spans the vector space of all length-$K$ signal trajectories that an LTI system can produce, provided that the collected input sequence is PE of order $K+n$ and the underlying system is controllable. However, the Fundamental Lemma requires a long continuous signal trajectory to construct the Hankel matrix. \cite{markovsky2022identifiability} extends the Fundamental Lemma to accommodate multiple short signal trajectories, which we refer to as discontinuous Fundamental Lemma in this paper. This extended Lemma is developed by using a general data structure called mosaic-Hankel matrix, which incorporates a dataset consisting of multiple short discontinuous signal trajectories.

\begin{definition}[Mosaic-Hankel Matrix] 
\label{Mosaic Hankel Matrix}
Let $W=\{w_{1:T_{1}}^{1}, \cdots, w_{1:T_{s}}^{s} \}$ be the set of $s$ discontinuous sequences with length of $T_{1}$, $\cdots$, $T_{s}$. The mosaic-Hankel matrix of depth $K \leq \min(T_1, \cdots, T_{s})$ is defined as:
\begin{equation}
  \begin{aligned}
    \label{Mosaic Hankel w}
    & \mathcal{M}_K(W) = [\mathcal{H}_K(w_{1:T_1}^{1}), \mathcal{H}_K(w_{1:T_2}^{2}), \cdots, \mathcal{H}_K(w_{1:T_s}^{s})].
  \end{aligned}
\end{equation}
Let $T=\sum_{i=1}^{s} T_{i}$ and $L=T-s(K-1)$, then we have $\mathcal{M}_K(W) \in \mathbb{R}^{qK \times L}$.
\end{definition}
\begin{lemma}[Discontinuous Fundamental Lemma \cite{markovsky2022identifiability}]
\label{Discontinuous Fundamental Lemma}
Consider the system \eqref{system} as a controllable LTI system with a pre-collected I/O sequence $U^{\mathrm{d}}=\{u_{1:T_1}^{\mathrm{d},1}, \cdots,  u_{1:T_s}^{\mathrm{d},s}\}$, $Y^{\mathrm{d}}=\{y_{1:T_1}^{\mathrm{d},1}, \cdots,  y_{1:T_s}^{\mathrm{d},s}\}$ of length $T$ which consists of $s$ I/0 sequences of length $T_{1}$, $\cdots$, $T_{s}$. Providing an input sequence $U^{\mathrm{d}}$ with
\begin{equation} \label{equ:rank}
r:= \mathrm{rank}\left( \begin{bmatrix} \mathcal{M}_{K}(U^{\mathrm{d}}) \\
\mathcal{M}_{K}(Y^{\mathrm{d}})
\end{bmatrix} \right) = mK+n,
\end{equation}
any length-$K$ sequence  $(u_{1:K}, y_{1:K})$ is an I/O trajectory of the LTI system if and only if we have
\begin{equation}
  \begin{aligned}
    \label{discontinuous fundamental lemma}
    \begin{bmatrix}
			u_{1:K} \\ y_{1:K}
		\end{bmatrix} = \begin{bmatrix}
			\mathcal{M}_{K}(U^{\mathrm{d}}) \\
			\mathcal{M}_{K}(Y^{\mathrm{d}})
		\end{bmatrix} g,
  \end{aligned}
\end{equation}
for some real vector $g \in \mathbb{R}^L$. 
\end{lemma}

It is worth noting that the Hankel matrix \eqref{Hankel w} represents a special case of the mosaic-Hankel matrix \eqref{Mosaic Hankel w} with $s=1$. The general form \eqref{Mosaic Hankel w} also describes other special forms, such as Page matrix or Trajectory matrix \cite{markovsky2021behavioral}. The main advantages of Lemma \ref{Discontinuous Fundamental Lemma} compared to Lemma \ref{Fundamental Lemma} are: i) it uses multiple short discontinuous trajectories instead of one long continuous trajectory, and ii) it replaces the PE condition on the input sequence with the rank condition~\eqref{equ:rank} on the I/O matrix. \eqref{equ:rank} is referred as generalized PE condition \cite{markovsky2022identifiability}. 

Both \eqref{fundamental lemma} and \eqref{discontinuous fundamental lemma} can be regarded as the non-parametric representation for system \eqref{system}. 
Let $T_{\mathrm{ini}}$, $N\in \mathbb{Z}$, and $K = T_{\mathrm{ini}} + N$. The Hankel~matrices $\mathcal{H}_{K}(u^{\mathrm{d}}_{1:T})$, $\mathcal{H}_{K}(y^{\mathrm{d}}_{1:T})$ (or mosaic-Hankel matrices $\mathcal{M}_{K}(U^{\mathrm{d}})$, $\mathcal{M}_{K}(Y^{\mathrm{d}})$) are divided into two parts (i.e., ``past data'' of length $T_{\mathrm{ini}}$ and ``future data'' of length $N$):
\begin{equation}\label{eq:Hankel}
\begin{aligned}
	\begin{bmatrix}U_{P}\\{U}_{F}
	\end{bmatrix} = \mathcal{H}_{K}({u^{\mathrm{d}}_{1:T}}), \quad
	\begin{bmatrix}{Y}_{P}\\{Y}_{F}
	\end{bmatrix} = \mathcal{H}_{K}({y^{\mathrm{d}}_{1:T}}),
\end{aligned}
\end{equation}
or 
\begin{equation}\label{eq:mosaic-Hankel}
\begin{aligned}
	\begin{bmatrix}U_{P}\\{U}_{F}
	\end{bmatrix} = \mathcal{M}_{K}(U^{\mathrm{d}}), \quad
	\begin{bmatrix}{Y}_{P}\\{Y}_{F}
	\end{bmatrix} = \mathcal{M}_{K}(Y^{\mathrm{d}}),
\end{aligned}
\end{equation}
where $U_{\mathrm{p}}$ and $U_{\mathrm{f}}$ denote the first $T_{\mathrm{ini}}$ block rows and the last $N$ block rows of $\mathcal{H}_{K}({u^{\mathrm{d}}_{1:T}})$ (or $\mathcal{M}_{K}(U^{\mathrm{d}})$), respectively (similarly for $Y_{\mathrm{p}}$ and $Y_{\mathrm{f}}$).
The data-enabled predictive control (DeePC) is formulated as \cite{coulson2019data,huang2023robust}
\begin{equation}
\centering
  \begin{aligned}
    \label{DeePC}
    &(y^{*}, u^{*}, \sigma_y^{*}, \sigma_u^{*}, g^{*}) = \underset{y, u, \sigma_y, \sigma_u, g}{\arg\min} \hspace{0 mm} J(y, u, \sigma_y,\sigma_u, g)\\
    &\mathrm{s.t.} \hspace{5 mm} \begin{bmatrix}
      U_{P}\\ U_F\\ Y_P\\ Y_F 
     \end{bmatrix} g = 
     \begin{bmatrix}
      u_{\mathrm{ini}}\\ u\\ y_{\mathrm{ini}}\\  y 
     \end{bmatrix} +
     \begin{bmatrix}
      \sigma_u\\ 0\\ \sigma_y\\ 0 
     \end{bmatrix},\\
    &\hspace{11 mm} u\in \mathcal{U}, y\in \mathcal{Y}.
  \end{aligned}
\end{equation}
In~\eqref{DeePC}, $J(y, u, \sigma_y,\sigma_u, g)$ represents the cost function. $u_{\mathrm{ini}}=u_{k-T_{\mathrm{ini}}:k-1}$ is the control input sequence within a past time horizon of length $T_{\mathrm{ini}}$, and $u= u_{k:k+N-1}$ is the control input sequence within a prediction horizon of length $N$ (similarly for $y_\mathrm{ini}$ and $y$). $\mathcal{U}$, $\mathcal{Y}$ represent the input and output constraints, respectively. $\sigma_{u}\in \mathbb{R}^{mT_{\mathrm{ini}}}$, $\sigma_{y} \in \mathbb{R}^{pT_{\mathrm{ini}}}$ stand for auxiliary variables.

Both Lemma \ref{Fundamental Lemma} and Lemma \ref{Discontinuous Fundamental Lemma} are only valid for deterministic LTI systems. \cite{berberich2022linear} and \cite{baros2022online} respectively show that Lemma \ref{Fundamental Lemma} can be extended to the nonlinear systems and the time-varying systems by continuously updating the Hankel matrix, which require a real-time PE input sequence. Moreover, \cite{teutsch2023online} removes the real-time PE requirement for the time-varying systems by using Lemma \ref{Discontinuous Fundamental Lemma}. However, the proposed rank condition (or PE condition) is not sufficient to ensure the informativeness of the data for non-deterministic LTI systems. While the rank condition may hold for the non-deterministic LTI systems, it can still lead to large prediction errors and poor control performance. In this paper, we propose an online DeePC framework for the time-varying systems to improve the rank condition and present a valid indicator for evaluating data informativeness.


\section{Main Result}
In this section, based on the data informativeness of the mosaic-Hankel matrix, we propose an online DeePC framework for the time-varying systems. The data informativeness is evaluated with the minimum non-zero singular value of the mosaic-Hankel matrix and is enhanced by adding the most informative signals to the matrix. It should be mentioned that the rank condition is always satisfied in real time since we add signal trajectories as additional columns. Moreover, we develop an online reduced-order DeePC using a numerical singular value decomposition (SVD) to reduce the computational cost of the control scheme.

\subsection{Online Data-Enabled Predictive Control}
For the deterministic LTI systems, Lemma \ref{Fundamental Lemma} and Lemma \ref{Discontinuous Fundamental Lemma} are valid only if the data is sufficiently informative. 
However, for the non-deterministic LTI systems, the collected data may result in a full-rank mosaic-Hankel matrix (or Hankel matrix) but cannot ensure accurate prediction of the system behavior.
To address this issue, \cite{coulson2022quantitative} proposes a quantitative measure of PE for the input trajectory based on the minimum non-zero singular value of the Hankel matrix, enhancing the robustness of the Fundamental Lemma against uncertainties. However, this approach is only effective locally around collected I/O data and leads to poor performance for the time-varying systems. Therefore, we focus on improving the performance of the discontinuous Fundamental Lemma for the time-varying systems. Singular values of a matrix are defined as $\sigma_1 \geq \sigma_2 \geq \cdots \geq \sigma_r \geq \cdots \geq \sigma_{\mathrm{end}} \geq 0$, where $\sigma_1$ and $\sigma_{\mathrm{end}}$ are also referred as $\sigma_{\mathrm{max}}$ and $\sigma_{\mathrm{min}}$, respectively. The singular value $\sigma_r$ corresponds to the rank of the matrix and is the minimum non-zero singular value, i.e., all singular values from $\sigma_{r+1}$ to $\sigma_{\mathrm{min}}$ are zero.

\cite{coulson2022quantitative} shows that the prediction error of the Fundamental Lemma can be arbitrarily large, even if the rank condition on the Hankel matrix is met. The important factor is the data informativeness, which is represented by the minimum non-zero singular value $\sigma_r$. Therefore, we can update the mosaic-Hankel matrix with real-time signal trajectories if $\sigma_r$ increases. If the real-time signal trajectory contains new information relevant to describing the system behavior \eqref{system}, it increases $\sigma_r$; otherwise, $\sigma_r$ decreases if the real-time signal trajectory lacks new information. Thus, we can set a threshold $\sigma_{\mathrm{thr}}$ for $\sigma_r$ to ensure that only the most informative data is added provided $\sigma_r \geq \sigma_{\mathrm{thr}}$. To achieve better prediction for the time-varying systems using the Fundamental Lemma, we formulate an online DeePC as follows:
\begin{equation}
\centering
  \begin{aligned}
    \label{online DeePC}
    &(y^{*}, u^{*}, \sigma_y^{*}, \sigma_u^{*}, g^{*}) = \underset{y, u, \sigma_y, \sigma_u, g}{\arg\min} \hspace{0 mm} J(y, u, \sigma_y, \sigma_u, g)\\
    &\mathrm{s.t.} \hspace{5 mm} \begin{bmatrix}
      U_P^o\\ U_F^o\\ Y_P^o\\ Y_F^o 
     \end{bmatrix} g = 
     \begin{bmatrix}
      u_{\mathrm{ini}}\\ u\\ y_{\mathrm{ini}}\\  y 
     \end{bmatrix} +
     \begin{bmatrix}
      \sigma_u\\ 0\\ \sigma_y\\  0 
     \end{bmatrix},\\
     &\hspace{11 mm} u\in \mathcal{U}, y\in \mathcal{Y},
  \end{aligned}
\end{equation}
where the matrices $U_P^o$, $U_F^o$, $Y_P^o$, and $Y_F^o$ are updated online. 
Specifically, denote $W^{\mathrm{d}}$ as the combined data set of $U^{\mathrm{d}}$ and $Y^{\mathrm{d}}$, i.e., $W^{\mathrm{d}}=\{ U^{\mathrm{d}}, Y^{\mathrm{d}} \}$. The corresponding mosaic-Hankel matrix $\mathcal{M}_{K}(W^{\mathrm{d}})$ is defined as:
\begin{equation} \label{eqn:Mw}
    \mathcal{M}_{K}(W^{\mathrm{d}}):= \begin{bmatrix}
		\mathcal{M}_{K}(U^{\mathrm{d}}) \\
		\mathcal{M}_{K}(Y^{\mathrm{d}})
	\end{bmatrix}.
\end{equation}
When $k< K$, the matrices $U_P^o$, $U_F^o$, $Y_P^o$, and $Y_F^o$ are initialized with $\mathcal{M}_{K}(W^{\mathrm{d}})$ (i.e., $U^{\mathrm{d}}$ and $Y^{\mathrm{d}}$ as shown in~\eqref{eq:mosaic-Hankel}). When $k\ge K$, the most recent real-time I/O sequence $w_{k-K+1:k}:=\begin{bmatrix}
    u_{k-K+1:k}^{\top}, y_{k-K+1:k}^{\top}
\end{bmatrix}^{\top}$ is first used to update the matrix
\[
\mathcal{M}_{K}(W^{\mathrm{d}}) \leftarrow \begin{bmatrix}
    \mathcal{M}_{K}(W^{\mathrm{d}}), w_{k-K+1:k}
\end{bmatrix}.
\]
Then, the rank and the minimum non-zero singular value of $\mathcal{M}_{K}(W^{\mathrm{d}})$ are calculated. If the minimum non-zero singular value of $\mathcal{M}_{K}(W^{\mathrm{d}})$ is larger than the threshold $\sigma_{\mathrm{thr}}$, i.e., $\sigma_r \geq \sigma_{\mathrm{thr}}$, then the matrices $U_P^o$, $U_F^o$, $Y_P^o$, and $Y_F^o$ are updated with $\mathcal{M}_{K}(W^{\mathrm{d}})$; otherwise, the sequence $w_{k-K+1:k}$ is removed from $\mathcal{M}_{K}(W^{\mathrm{d}})$.

The proposed online DeePC framework is summarized in Algorithm~\ref{ODeePC}. 
A drawback of Algorithm~\ref{ODeePC} is that the dimension of the mosaic-Hankel matrix $\mathcal{M}_{K}(W^\mathrm{d})$ grows by adding real-time data, leading to high computational cost for the online DeePC. To address this challenge, we next propose an order reduction scheme to improve the computational efficiency of online DeePC.

\begin{figure}
\removelatexerror
\scalebox{0.85}{
\begin{algorithm*}[H]
\SetAlFnt{\small}
    \SetKwInOut{Parameter}{Parameter}
    \SetKwInOut{Input}{Input}
    \SetKwInOut{Output}{Output}
\caption{Online Data-Enabled Predictive Control}
\label{ODeePC}
\SetAlgoLined
\Input{$T_{\mathrm{ini}}$, $N$, $T_{c}$, $\sigma_{\mathrm{thr}}$, $J(y,u,\sigma_y,\sigma_u,g)$, $\{\mathcal{U}, \mathcal{Y}\}$, $\{U^{d}, Y^{d}\}$.}
\Output{$u_{1:T_c}$, $y_{1:T_c}$.}
\vspace{0.2em}
\hrule
\vspace{0.2em}
Generate $\mathcal{M}_{K}(U^{\mathrm{d}})$ and $\mathcal{M}_{K}(Y^{\mathrm{d}})$\;
Initialize $\{U_P^o, Y_P^o, U_F^o, Y_F^o\}$ and $\mathcal{M}_K(W^d)$\;
Initialize $u_{\mathrm{ini}}=u_{1:T_{\mathrm{ini}}}$, $y_{\mathrm{ini}}=y_{1:T_{\mathrm{ini}}}$\;
\For{$k = T_{\mathrm{ini}}$ : $T_c$}{
     Solve optimization problem \eqref{online DeePC} to obtain $u^*$\;
     Apply first control input, i.e., $u^*(1)$, to the plant\;
     Measure $y(k)$ from the plant\;
     Update $u_{\mathrm{ini}}=u_{k-T_{\mathrm{ini}}+1:k}$ and $y_{\mathrm{ini}}=y_{k-T_{\mathrm{ini}}+1:k}$\;
     \If{$k \ge K$}{
        Generate $w_{k-K+1:k}$\;
        Update $\mathcal{M}_K(W^{\mathrm{d}}) \leftarrow [\mathcal{M}_K(W^\mathrm{d}), w_{k-K+1:k}]$\;
        Calculate $r = \mathrm{rank}(\mathcal{M}_K(W^{\mathrm{d}}))$\;
        Calculate $\sigma_{r}(\mathcal{M}_K(W^{\mathrm{d}}))$\;
        \If{$\sigma_r(\mathcal{M}_K(W^{\mathrm{d}})) \geq \sigma_{\mathrm{thr}}$}{
        Update $\{U_P^o, Y_P^o, U_F^o, Y_F^o\}$\;
     }
     \Else{
        Remove signal $w_{k-K+1:k}$ from $\mathcal{M}_K(W^{\mathrm{d}})$\;
     }
     }     
}
\end{algorithm*}}
\end{figure}

\subsection{Online Reduced-Order Data-Enabled Predictive Control}
SVD techniques are effective in reducing computational complexity for data-driven control methods~\cite{Zhang2023,shi2023efficient}. In this section, we incorporate a new numerical SVD algorithm into the online DeePC framework such that the reduced-order mosaic-Hankel matrix and corresponding singular values can be updated efficiently.

Considering $r = \mathrm{rank}(\mathcal{M}_K(W^\mathrm{d}))$, one can formulate the SVD of the mosaic-Hankel matrix $\mathcal{M}_K(W^{\mathrm{d}})$ as follows:
\begin{equation}
  \begin{aligned}
    \label{SVD M}
    &\mathcal{M}_K(W^\mathrm{d}) = U \Sigma V^{\top}= [U_r \hspace{2 mm} U_{qK-r}] \begin{bmatrix}
    \Sigma_r & 0\\
    0 & 0
    \end{bmatrix} [V_r \hspace{2 mm} V_{L-r}]^{\top},
  \end{aligned}
\end{equation}
where $q=m+p$, $\Sigma \in \mathbb{R}^{qK \times L}$ is the singular matrix, and $U \in \mathbb{R}^{qK \times qK}$ and $V \in \mathbb{R}^{L \times L}$ are left and right singular vectors, respectively, such that $UU^{\top}=U^{\top}U=I_{qK}$ and $VV^{\top}=V^{\top}V=I_L$. Moreover, $\Sigma_r$ contains the top $r$ non-zero singular values, $U_r \in \mathbb{R}^{qK \times r}$, $U_{qK-r} \in \mathbb{R}^{qK \times (qK-r)}$, and $V_r \in \mathbb{R}^{L \times r}$, $V_{L-r} \in \mathbb{R}^{L \times (L-r)}$. 
Therefore, one can write
\begin{equation}
  \begin{aligned}
    \label{reduced M}
    &\mathcal{M}_K(W^\mathrm{d}) g = U_r \Sigma_r V_r^{\top} g = \mathcal{M}_K'(W^{\mathrm{d}}) g',
  \end{aligned}
\end{equation}
where $\mathcal{M}_K'(W^\mathrm{d}) = U_r \Sigma_r \in \mathbb{R}^{qK \times r}$ and $g' = V_r^{\top} g \in \mathbb{R}^r$. If the pre-collected data is sufficient rich, then we have $mK+n \leq r \leq \mathrm{min}(qK, L)$. Thus, one can approximate \eqref{reduced M} using a rank order $mK+n \leq r_a \leq r$, as follows:
\begin{equation}
  \begin{aligned}
    \label{minimum M}
    &\mathcal{M}_K(W^\mathrm{d}) g \approx U_{r_a} \Sigma_{r_a} V_{r_a}^{\top} g = \mathcal{M}_K''(W^\mathrm{d}) g'',
  \end{aligned}
\end{equation}
where $\mathcal{M}_K''(W^\mathrm{d}) = U_{r_a} \Sigma_{r_a} \in \mathbb{R}^{qK \times r_a}$ and $g'' = V_{r_a}^{\top} g \in \mathbb{R}^{r_a}$. 

Now, one can formulate an online reduced-order DeePC as follows:
\begin{equation}
\centering
  \begin{aligned}
    \label{online reduced-DeePC}
    &(y^{*}, u^{*}, \sigma_y^{*}, \sigma_u^{*},{g''}^{*}) = \underset{y, u, \sigma_y, \sigma_u, {g''}}{\arg\min} \hspace{0 mm} J({y}, u, \sigma_y, \sigma_u, {g''})\\
    &\mathrm{s.t.} \hspace{5 mm} \begin{bmatrix}
      U^{o''}_P\\ U_F^{o''}\\ Y_{P}^{o''}\\ Y_F^{o''} 
     \end{bmatrix} g'' = 
     \begin{bmatrix}
      u_{\mathrm{ini}}\\ u\\ y_{\mathrm{ini}}\\  y 
     \end{bmatrix} +
     \begin{bmatrix}
      \sigma_u\\ 0\\ \sigma_y\\  0 
     \end{bmatrix},\\
     &\hspace{11 mm} u\in \mathcal{U}, y\in \mathcal{Y},
  \end{aligned}
\end{equation}
where the matrices $U^{o''}_P$, $U_F^{o''}$, $Y_P^{o''}$, and $Y_F^{o''}$ are updated online based on $\mathcal{M}_K''(W^\mathrm{d})$ under an adaptive order $r_a$ such that $\sigma_{r_a} \geq \sigma_{\mathrm{thr}}$. It should be mentioned that we use SVD to update $\mathcal{M}_K''(W^\mathrm{d})$ for $\mathcal{M}_{K}(W^{\mathrm{d}}) \leftarrow \begin{bmatrix} \mathcal{M}_{K}(W^{\mathrm{d}}), w_{k-K+1:k} \end{bmatrix}$. Moreover, the adaptive order $mK+n \leq r_a \leq r$, which is based on the threshold singular value $\sigma_{\mathrm{thr}}$, allows an adaptive dimension for the reduced-order mosaic-Hankel matrix $\mathcal{M}_K''(W^\mathrm{d})$ regarding the data informativity. Indeed, the dimension of $\mathcal{M}_K''(W^{\mathrm{d}})$, i.e., $r_a$, is changed adaptively based on $\sigma_{r_a} \geq \sigma_{\mathrm{thr}}$. 


For both proposed online DeePC frameworks, calculating the SVD at each time step is not computationally efficient for real-time control. Therefore, inspired by \cite{brand2006fast}, we propose a numerical algorithm to compute the SVD of $\begin{bmatrix} \mathcal{M}_{K}(W^{\mathrm{d}}), w_{k-K+1:k} \end{bmatrix}$ by taking advantage of our knowledge of the SVD of $\mathcal{M}_K(W^{\mathrm{d}})$, which reduces the computational time of the proposed online DeePC. 

When $\mathrm{rank}(\mathcal{M}_K(W^\mathrm{d}))< \mathrm{rows}(\mathcal{M}_K(W^\mathrm{d}))$, one can express $\begin{bmatrix} \mathcal{M}_{K}(W^{\mathrm{d}}), w_{k-K+1:k} \end{bmatrix}$ as $X+AB^{\top}$, where $X = [\mathcal{M}_K(W^{\mathrm{d}}), 0]$, $A=w_{k-K+1:k}$, and $B=[0, \cdots, 0, 1]^{\top}$. Therefore, one has
\begin{equation}
  \begin{aligned}
    \label{SVD Adding 1}
    &X+AB^{\top} = [U_r \hspace{2 mm} A] \begin{bmatrix}
    \Sigma_r & 0\\
    0 & I
    \end{bmatrix} [V_x \hspace{2 mm} B]^{\top}.
  \end{aligned}
\end{equation}
where $V_x = [V_r^{\top} \hspace{2 mm} 0]^{\top}$.
Let $P$ be an orthogonal basis of the column space of $(I-U_r U^{\top}_r)A$, i.e., the component of $A$ that is orthogonal to $U_r$, and set $R_A = P^{\top}(I-U_r U^{\top}_r)A$. Now, one can write
\begin{equation}
  \begin{aligned}
    \label{SVD Adding 2}
    &[U_r \hspace{2 mm} A] = [U_r \hspace{2 mm} P] \begin{bmatrix}
    I & U^{\top}_r A\\
    0 & R_A
    \end{bmatrix},
  \end{aligned}
\end{equation}
where similar to a QR decomposition, $R_A$ needs not be upper-triangular or square. Similarly, let $R_B = Q^{\top}(I-V_x V_x^{\top})B$, where $Q$ is the component of $B$ that is orthogonal to $V_x$. Thus, one has
\begin{equation}
  \begin{aligned}
    \label{SVD Adding 3}
    &[V_x \hspace{2 mm} B] = [V_x \hspace{2 mm} Q] \begin{bmatrix}
    I & V^{\top}_x B\\
    0 & R_B
    \end{bmatrix}.
  \end{aligned}
\end{equation}
Substituting \eqref{SVD Adding 2} and \eqref{SVD Adding 3} into \eqref{SVD Adding 1}, we have
\begin{equation}
  \begin{aligned}
    \label{SVD Adding 4}
    &X+AB^{\top} = [U_r \hspace{2 mm} P] S [V_x \hspace{2 mm} Q]^{\top},\\ 
    &S = \begin{bmatrix}
    I & U^{\top}_r A\\
    0 & R_A
    \end{bmatrix} \begin{bmatrix}
    \Sigma_r & 0\\
    0 & I
    \end{bmatrix} \begin{bmatrix}
    I & V^{\top}_x B\\
    0 & R_B
    \end{bmatrix}^{\top},
  \end{aligned}
\end{equation}
where one can write $S$ as:
\begin{equation}
  \begin{aligned}
    \label{SVD Adding 5}
    &S = \begin{bmatrix}
    \Sigma_r & 0\\
    0 & 0
    \end{bmatrix} + \begin{bmatrix}
    U^{\top}_r A\\
    R_A
    \end{bmatrix} \begin{bmatrix}
    V^{\top}_x B\\
    R_B
    \end{bmatrix}^{\top}.
  \end{aligned}
\end{equation}
Using \cite{gu1993stable}, diagonalizing $S = U_S \Sigma_S V_S^{\top}$ gives rotations $U_S$ and $V_S$ of the extended subspaces $[U_r \hspace{2 mm} P]$ and $[V_x \hspace{2 mm} Q]$ such that
\begin{equation}
  \begin{aligned}
    \label{SVD Adding 6}
    &X+AB^{\top} = ([U_r \hspace{2 mm} P]U_S) \Sigma_S ([V_x \hspace{2 mm} Q]V_S)^{\top}\\ 
  \end{aligned}
\end{equation}
is the desired SVD.

\begin{remark} [Rank-1 Modifications]
\label{Rank-1 modifications}
For the proposed numerical SVD, we are not limited to only add one signal to the mosaic-Hankel matrix. The above formulations work for adding more signals at the same time by defining the correct matrices $A$ and $B$. However, in our algorithm, we focus on adding one signal to the matrix at each time step and calculate the SVD of the new matrix based on the SVD of the original matrix, which is called Rank-1 modification in the numerical SVD. For Rank-1 modification, we define $P = \lVert (I-U_r U_r^T)A \rVert^{-1} (I-U_r U_r^T)A$ and $Q = \lVert (I-V_x V_x^T)B \rVert^{-1} (I-V_x V_x^T)B$, which yield $S = \begin{bmatrix} \Sigma_r & U_r^TA\\ 0 & R \end{bmatrix}$ since $V_x^TB = [V_r^T \hspace{2 mm} 0] [0,...,0,1]^T = 0$. 
\end{remark}

When $\mathrm{rank}(\mathcal{M}_K(W^\mathrm{d}))= \mathrm{rows}(\mathcal{M}_K(W^\mathrm{d}))$, one needs rewrite the SVD of the mosaic-Hankel matrix $\begin{bmatrix} \mathcal{M}_{K}(W^{\mathrm{d}}), w_{k-K+1:k} \end{bmatrix}$ as
\begin{equation}
  \begin{aligned}
    \label{SVD Adding 20}
    &X+AB^{\top} = U_r [\Sigma_r \hspace{2 mm} U^{\top}_r A] [V_x \hspace{2 mm} B]^{\top}
  \end{aligned}
\end{equation}
Therefore, we only need to provide an orthogonal matrix for $[V_x \hspace{2 mm} B]$, which yields 
$S = \begin{bmatrix} \Sigma_r & 0 \end{bmatrix} + U^{\top}_r A \begin{bmatrix} V^{\top}_x B\\ R_B \end{bmatrix}^{\top}$, and $X+AB^{\top} = (U_rU_S) \Sigma_S ([V_x \hspace{2 mm} Q]V_S)^{\top}$ is the desired SVD.

The proposed online reduced-order DeePC is presented in Algorithm 2, which uses the above modified version of the numerical SVD \cite{brand2006fast}. Moreover, Algorithm 2 includes finding the adaptive order $r_a$ for the reduced-order mosaic hankel matrix $\mathcal{M}_K''(W^{\mathrm{d}})$ regarding the data informativity, which is shown in Algorithm 3.

\begin{remark} [Comparison]
\label{Comparison}
Compared to \cite{shi2023efficient}, the proposed online reduced-order DeePC measures data informativity of real-time signals by observing the minimum non-zero singular value of the mosaic-Hankel matrix, selectively using the most informative signals instead of adding all real-time trajectories. Moreover, the dimension of the online reduced-order DeePC starts from $r_a = mK+n$ instead of the rank, which leads to the minimum possible dimension for the mosaic-Hankel matrix. 
We also refine the numerical SVD algorithm, addressing its speed and stability issues as noted in \cite{gu1993stable}.
These three contributions improve the control performance and reduce computational complexity.
\end{remark}

\begin{figure}
\removelatexerror
\scalebox{0.85}{
\begin{algorithm*}[H]
\SetAlFnt{\small}
    \SetKwInOut{Parameter}{Parameter}
    \SetKwInOut{Input}{Input}
    \SetKwInOut{Output}{Output}
\caption{Online Reduced-Order DeePC}
\label{ORODeePC}
\SetAlgoLined
\Input{$T_{\mathrm{ini}}$, $N$, $T_{c}$, $\sigma_{\mathrm{thr}}$, $J(y,u,\sigma_y,\sigma_u,g)$, $\{\mathcal{U}, \mathcal{Y}\}$, $\{U^{d}, Y^{d}\}$.}
\Output{$u_{1:T_c}$, $y_{1:T_c}$.}
\vspace{0.2em}
\hrule
\vspace{0.2em}
Generate $\mathcal{M}_{K}(U^{\mathrm{d}})$ and $\mathcal{M}_{K}(Y^{\mathrm{d}})$\;
Generate $\{U_P^o, Y_P^o, U_F^o, Y_F^o\}$ and $\mathcal{M}_K(W^d)$\;
Calculate $r = \mathrm{rank}(\mathcal{M}_K(W^d))$ and $\mathcal{M}_K'(W^d)$\;
Calculate $r_a=mK+n$ and $\mathcal{M}_K''(W^d)$\;
Initialize $\{U_P^{o''}, Y_P^{o''}, U_F^{o''}, Y_F^{o''}\}$\;
Initialize $u_{\mathrm{ini}}=u_{1:T_{\mathrm{ini}}}$, $y_{\mathrm{ini}}=y_{1:T_{\mathrm{ini}}}$\;
\For{$k = T_{ini}$ : $T_c$}{
     Solve optimization problem \eqref{online reduced-DeePC} to obtain $u^*$\;
     Apply first control input, i.e., $u^*(1)$, to the plant\;
     Measure $y(k)$ from the plant\;
     Update $u_{\mathrm{ini}}=u_{k-T_{\mathrm{ini}}+1:k}$ and $y_{\mathrm{ini}}=y_{k-T_{\mathrm{ini}}+1:k}$\;
     \If{$k >= K$}{
        Generate $w_{k-K+1:k}$\;
        \If{$r < qK$}{
        Calculate $P = \lVert (I-U_r U_r^{\top}) w_{k-K+1:k} \rVert^{-1}$\;
        Calculate $R = P^T (I-U_r U_r^T) w_{k-K+1:k}$\;
        Calculate $S = \begin{bmatrix}
        \Sigma_r & U_r^T w_{k-K+1:k}\\
        0 & R
        \end{bmatrix}$\;
        Calculate numerical SVD $S \rightarrow U_S \Sigma_S V_S^T$\;
        \If{$rank(\Sigma_S) == r+1$}{
        Update $U_r = [U_r \hspace{1mm} P] U_S$, $\Sigma_r = \Sigma_S$\;
        Update $r = r+1$\;
        }
        \Else{
        Update $U_r = [U_r \hspace{1mm} P] U_{S_r}$, $\Sigma_r = \Sigma_{S_r}$\;
        }
        }
        \Else{
        Calculate $S = \begin{bmatrix}
        \Sigma_r & U_r^{\top} w_{k-K+1:k}
        \end{bmatrix}$\;
        Calculate numerical SVD $S \rightarrow U_S \Sigma_S V_S^{\top}$\;
        Update $U_r = U_r U_S$, $\Sigma_r = \Sigma_{S_r}$\;
        }
        Calculate adaptive order $r_a$\ (See Algorithm 3)\;
        Update $\mathcal{M}_K^{''}(W^d) = U_{r_a} \Sigma_{r_a}$\;
        Update $\{U^{o''}_P, Y^{o''}_P, U^{o''}_F, Y^{o''}_F\}$\;
     }     
}
\end{algorithm*}}
\end{figure}

\begin{figure}
\removelatexerror
\scalebox{0.85}{
\begin{algorithm*}[H]
\SetAlFnt{\small}
    \SetKwInOut{Parameter}{Parameter}
    \SetKwInOut{Input}{Input}
    \SetKwInOut{Output}{Output}
\caption{Adaptive Rank Order $r_a$}
\label{Adaptive $r_a$}
\SetAlgoLined
\Input{$\Sigma_r$, $\sigma_{thr}$.}
\Output{$r_a$.}
\vspace{0.2em}
\hrule
\vspace{0.2em}
\For{$i = mK+n:r$}{
     \If{$\sigma_i >= \sigma_{thr}$}{
        Update $r_a = i$\;
     }
     \Else{
        Break\;
     }    
}
\If{$\sigma_{mK+n} < \sigma_{thr}$}{
        Update $r_a = mK+n$\;
}
\end{algorithm*}}
\end{figure}

\section{Simulation Results}
\subsection{Linear Time-Varying System}
In this subsection, we demonstrate the performance of the proposed online reduced-order DeePC framework via a simulation example on a linear time-varying (LTV) system, described by: 
\begin{equation}
  \begin{aligned}
    \label{LTV system}
    x(k+1) &= A(k)x(k)+B(k)u(k)+d_p(k),\\ 
    y(k) &= Cx(k)+d_m(k),
  \end{aligned}
\end{equation}
where $d_p$ and $d_m$ represent process and measurement noises, respectively. The matrices $A(k)$, $B(k)$, and $C$ are constructed as 
\begin{equation*}
  \begin{aligned}
  & A(k) =A^o + \lambda(k) \begin{bmatrix}
    0.01 & 0 & 0.001 & 0\\
    0 & 0.01 & 0 & 0.001\\
    0 & 0 & 0.01 & 0\\
    0 & 0 & 0 & 0.01\\
    \end{bmatrix},\\
 & B(k) = B^o + \lambda(k) \begin{bmatrix}
    0.001 & 0.0001\\
    0.0001 & 0.001\\
    0 & 0.001\\
    0.001 & 0\\
    \end{bmatrix},\\
 & C = \begin{bmatrix}
    1 & 0 & 0 & 0\\
    0 & 1 & 0 & 0\\
    \end{bmatrix},
  \end{aligned}
\end{equation*}
where $\lambda(k)$ is a time-varying parameter, and $A^o$ and $B^o$ are set as follows:
\begin{equation*}
  \begin{aligned}
  & A^o = \begin{bmatrix}
    0.921 & 0 & 0.041 & 0\\
    0 & 0.918 & 0 & 0.033\\
    0 & 0 & 0.924 & 0\\
    0 & 0 & 0 & 0.937\\
    \end{bmatrix},\\
 & B^o = \begin{bmatrix}
    0.017 & 0.001\\
    0.001 & 0.023\\
    0 & 0.061\\
    0.072 & 0\\
    \end{bmatrix}.
  \end{aligned}
\end{equation*}
For the simulation, we consider the following values: $T_{\mathrm{ini}}=35$, $N = 45$, the simulation time $T_c = 2100$, $x(0) = [0.5,0.5,0.5,0.5]^{\top}$, and $d_p$ and $d_m$ are considered random values such that $\lVert d_p \rVert,\lVert d_m \rVert \leq 0.002$. Moreover, the initial trajectory $(u_{\mathrm{ini}},y_{\mathrm{ini}})$ is generated by applying zero control input to the system and measuring the corresponding output. 

\begin{figure}[!h]
     \centering
     \includegraphics[width=0.99\linewidth]{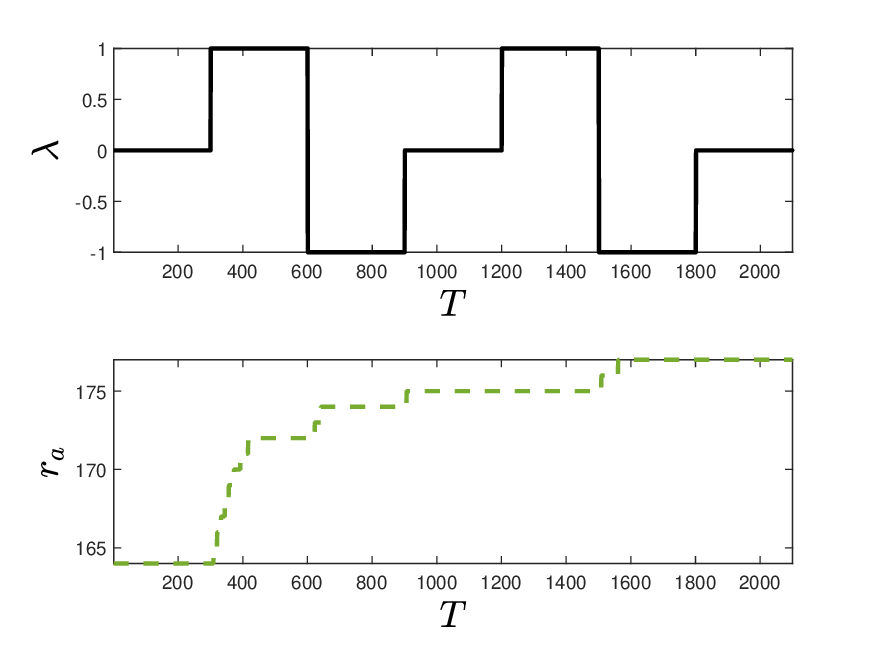}
    \caption{Order of reduced-order mosaic-Hankel matrix for LTV system.}
     \label{Adaptive Order}
 \end{figure}

The evolution of $\lambda(k)$ is shown in Fig. \ref{Adaptive Order}, where one can see that the behavior of the LTV system is repeating in real time. Fig. \ref{Adaptive Order} also depicts the order of the reduced-order mosaic-Hankel matrix for the proposed online DeePC \eqref{online reduced-DeePC}, showing that the adaptive order $r_a$ changes when the LTV system switches to different dynamics. For the online DeePC \cite{shi2023efficient}, the order is constant and equal to the rank of the mosaic-Hankel matrix, which is $320$ for the collected data set.

\begin{figure}[!h]
     \centering
     \includegraphics[width=0.99\linewidth]{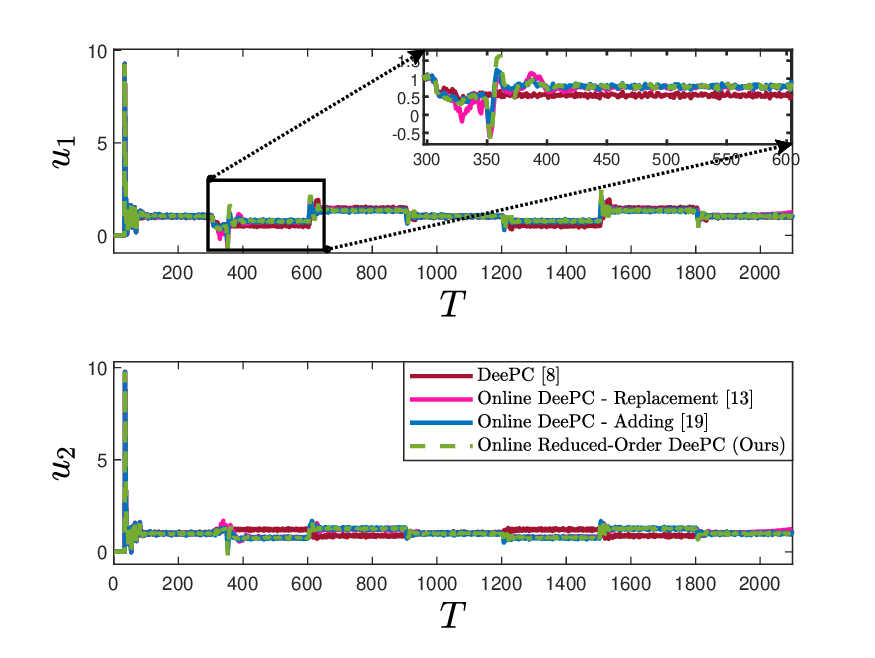}
    \caption{Comparison of control inputs for the LTV system.}
     \label{Inputs}
 \end{figure}

\begin{figure}[!h]
     \centering
     \includegraphics[width=0.99\linewidth]{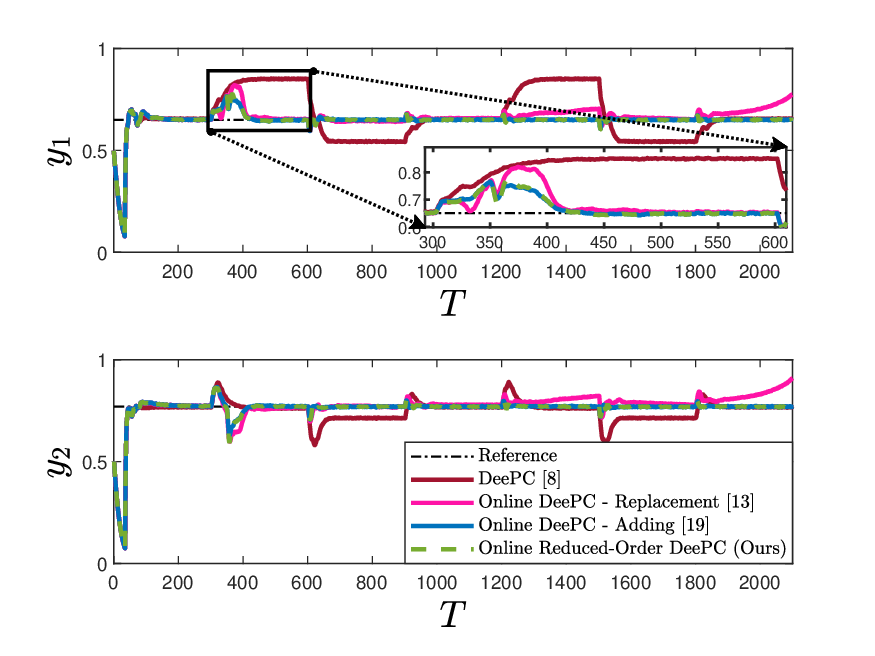}
    \caption{Comparison of system outputs of the LTV system.}
     \label{Outputs}
 \end{figure}

Figs. \ref{Inputs} and \ref{Outputs} show the control performance of the proposed online reduced-order DeePC \eqref{online reduced-DeePC} in comparison with the traditional DeePC \cite{coulson2019data}, the online DeePC \cite{teutsch2023online} (replacing old data with new data in the mosaic-Hankel matrix), and the online DeePC \cite{shi2023efficient} (adding new data to the mosaic-Hankel matrix). For the traditional DeePC \cite{coulson2019data}, the DeePC policy \eqref{DeePC} is employed such that the first optimal control input is applied to the system, and the initial trajectory $\{u_{\mathrm{ini}}, y_{\mathrm{ini}}\}$ is updated for the next step (see Algorithm 2 in \cite{coulson2019data}). From Figs. \ref{Inputs} and \ref{Outputs}, one can see that the proposed online DeePC shows better tracking performance in comparison with other control schemes since it uses informative data to update the mosaic-Hankel matrix. The online DeePC \cite{teutsch2023online} does not show a reasonable performance when all PE data is removed from the mosaic-Hankel matrix since the rank condition is not enough to evaluate suitable data. Table I demonstrates that the developed online DeePC significantly reduces the computational time compared to other controllers while keep the tracking performance well. Moreover, the average time for computing the numerical SVD is $0.0261 ms$ for the online DeePC \cite{shi2023efficient}; however, our proposed numerical SVD \eqref{SVD Adding 20} takes $0.0085 ms$, which leads to lower computational complexity for the the online DeePC \eqref{online reduced-DeePC}. The dimension of the online reduced-order DeePC \eqref{online reduced-DeePC} is another reason for better computational time since it is the minimum possible dimension for the mosaic-Hankel matrix. It is worth noting that Track. Perf. represents tracking performance, which is root mean squared error as $\lVert reference - y \rVert$, for different control frameworks. 


\begin{table}[!ht]
\centering
 \caption{Comparison of Tracking Performance and Computational Cost for LTV System}
\begin{tabular}{ |p{3.8cm}|p{1.8cm}|p{2.2cm}| }
\hline
\hline
Controller & Track. Perf. & Time (per loop) \\
\hline
DeePC [8] & $4.83$ & $0.21 \hspace{1 mm} s$ \\\hline
Online DeePC [13] & $3.39$ & $0.18 \hspace{1 mm} s$ \\\hline
Online DeePC [19] & $\bf{3.05}$ & $0.09 \hspace{1 mm} s$ \\\hline
Online R.O. DeePC  (Ours) & $\bf{3.05}$ & $\bf{0.05} \hspace{1 mm} s$ \\\hline
\hline
\end{tabular}
\end{table}

\subsection{Vehicle Rollover Avoidance}
Rollover is a type of vehicle accident in which a vehicle tips over onto its side or roof. The rollover propensity of a vehicle is changed for different road surfaces or carried loads. Therefore, it is a big challenge to derive an accurate model for the vehicle dynamics which includes all operating conditions. We apply our online reduced-order DeePC to safeguard a vehicle against rollover. Considering a constant longitudinal speed for the vehicle, the steering wheel angle (SW) acts as the command, which is generated either by a human operator or a higher-level planning algorithm. For an arbitrary reference command SW, denoted as $u_r$, the rollover constraint may not be satisfied. Thus, DeePC is used as a safety filter for the reference command SW $u_r$ to obtain an admissible input $u$. Indeed, DeePC ensures compliance with the load transfer ratio (LTR) constraint to prevent potential rollovers.

Following \cite{bencatel2017reference}, a linear model is considered to represent the vehicle dynamics, which vehicle roll angle $q$, roll rate $p$, lateral velocity $v$, and yaw rate $\gamma$ represent the states of the system. Considering the system in the format of \eqref{LTV system}, we have $A(k) = A^o + 0.01 \lambda(k) A^o$, $B(k) = B^o + 0.01 \lambda(k) B^o$ as the time-varying matrices, and $A^o$, $B^o$, and $C$ are considered as:
\begin{equation*}
  \begin{aligned}
  & A^o = T_s\begin{bmatrix}
    0.00499 & 0.997 & 0.0154 & -6.81 \times 10^{-5}\\
    -78.3 & -12.2 & -65.3 & -3.89\\
    -0.932 & -0.799 & -6.20 & -1.57\\
    1.52 & 3.32 & 8.27 & -1.49\\
    \end{bmatrix} + I_4,\\
 & B^o = T_s\begin{bmatrix}
    -5.76 \times 10^{-5} & 2.80 & 0.278 & 0.655\\
    \end{bmatrix}^T,\\
 & C = \begin{bmatrix}
    0.1200 & 0.0124 & -0.0108 & 0.0109\\
    \end{bmatrix},
  \end{aligned}
\end{equation*}
where $x = \begin{bmatrix} q & p & v & \gamma \end{bmatrix}^T$, $u = SW$, $y = LTR$, and $T_s$ is sampling time. It should be mentioned that the matrices $A^o$, $B^o$, and $C$ are obtained based on a CarSim model for a standard utility truck under a constant longitudinal speed $80 km/h$. More specifically, the vehicle tracks a constant reference longitudinal speed $80 km/h$ using a feedback control on the gas pedal, which is not discussed here.

Through the LTR, the rollover constraint is defined as:
\begin{equation*}
  \begin{aligned}
  & LTR = \frac{F_{z,R} - F_{z,L}}{mg},
  \end{aligned}
\end{equation*}
where $mg$ is the vehicle weight, and $F_{z,R}$ and $F_{z,L}$ stand for the total vertical force on the right-side tires and the left-side tires, respectively. Note that $|LTR| > 1$ means wheels lifting off; thus, the rollover constraint is imposed as:
\begin{equation*}
  \begin{aligned}
  & -1 \leq LTR \leq 1.
  \end{aligned}
\end{equation*}

\begin{figure}[!h]
     \centering
     \includegraphics[width=0.99\linewidth]{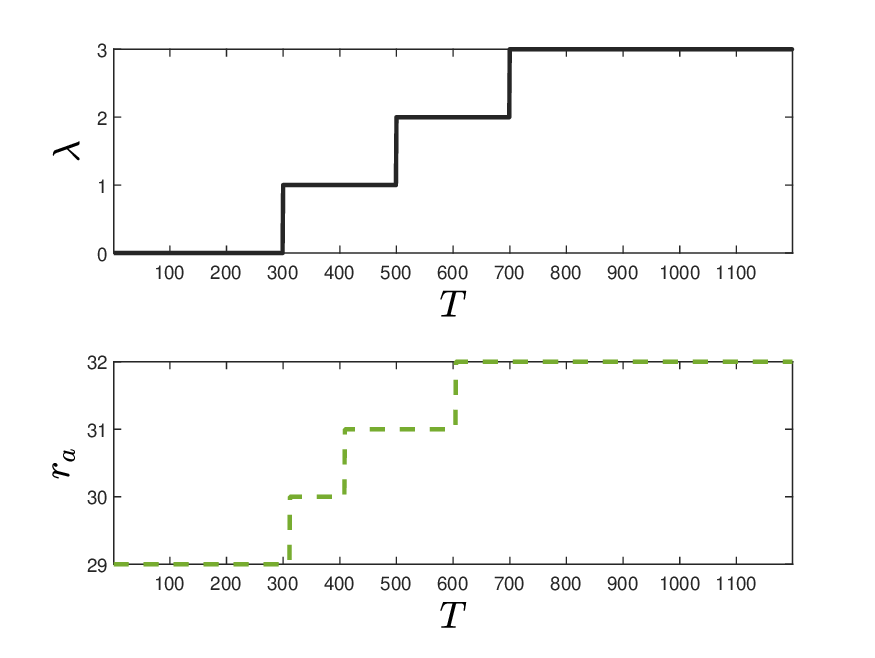}
    \caption{Order of reduced-order mosaic-Hankel matrix for vehicle rollover avoidance.}
     \label{Adaptive OrderV}
 \end{figure}

\begin{figure}[!h]
     \centering
     \includegraphics[width=0.99\linewidth]{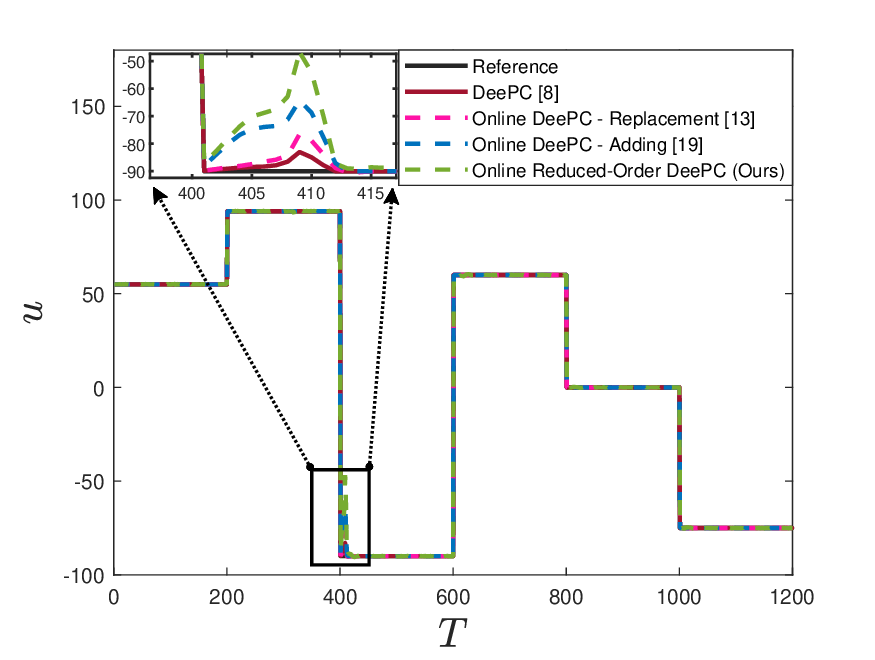}
    \caption{Control inputs for vehicle rollover avoidance.}
     \label{InputsV}
 \end{figure}

\begin{figure}[!h]
     \centering
     \includegraphics[width=0.99\linewidth]{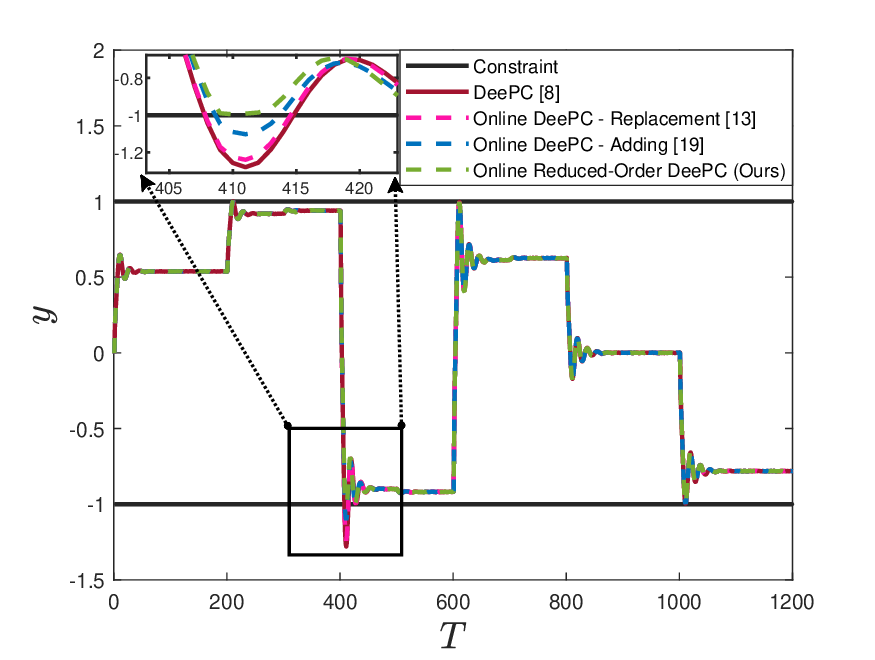}
    \caption{System outputs for vehicle rollover avoidance.}
     \label{OutputsV}
 \end{figure}

For the simulation, we consider the following values: $T_{ini}=10$, $N = 15$, the simulation time $T_c = 1200$, the smapling time $T_s = 0.1$, $x(0) = [0,0,0,0]^T$, and random values $\lVert d_p \rVert,\lVert d_m \rVert \leq 0.002$. Moreover, the first initial trajectory $(u_{ini},y_{ini})$ is generated by applying $u(k) = 55, k = 1:T_{ini}$ to the system and measuring the corresponding output. For this case, $\lambda(k)$ and $r_a$ are shown in Fig. \ref{Adaptive OrderV}, where one can see that the adaptive order $r_a$ changes when the system switches to another dynamics. For the online DeePC \cite{shi2023efficient}, the rank of the mosaic-Hankel matrix is $50$, which is the order of the reduced-order mosaic-Hankel matrix. Like the previous case, Fig. \ref{Adaptive OrderV} illustrates that the online DeePC \eqref{online reduced-DeePC} has lower order than the online DeePC \cite{shi2023efficient}, which leads to better computational cost. Figs. \ref{InputsV} and \ref{OutputsV} show the control performance of the proposed online reduced-order DeePC \eqref{online reduced-DeePC} in comparison with the traditional DeePC \cite{coulson2019data}, the online DeePC \cite{teutsch2023online} (replacing old data by new data in the mosaic-Hankel matrix), and the online DeePC \cite{shi2023efficient} (adding new data to the mosaic-Hankel matrix). Fig. \ref{InputsV} shows the tracking performance and modification to the reference command SW for the control strategies. As shown in Fig. \ref{OutputsV}, one can see that the proposed online reduced-order DeePC only satisfies the LTR constraint, and other control schemes cannot satisfy the constraint. Table II demonstrates that the developed online DeePC significantly reduces the computational time compared to other controllers while keep the tracking performance and system safety well. It should be mentioned that Const. Viol. represents constraint violation number for different control frameworks. In this case, not only our online DeePC has lower computational cost compared to the online DeePC \cite{shi2023efficient}, but also it only satisfies the safety constraint due to employing useful information in the mosaic-Hankel matrix.

\begin{table}[!ht]
\centering
 \caption{Comparison of Safety Performance and Computational Cost for vehicle rollover avoidance}
\begin{tabular}{ |p{3.8cm}|p{1.8cm}|p{2.2cm}|  }
\hline
\hline
Controller & Const. Viol. & Time (per loop) \\
\hline
DeePC [8] & 1 & $0.024 \hspace{1 mm} s$ \\\hline
Online DeePC [13] & 1 & $0.025 \hspace{1 mm} s$ \\\hline
Online DeePC [19] & 1 & $0.010 \hspace{1 mm} s$ \\\hline
Online R.O. DeePC  (Ours)  & \bf{0} & $\bf{0.007} \hspace{1 mm} s$ \\\hline
\hline
\end{tabular}
\end{table}

\section{Conclusion}
\label{Sec6}
This paper proposed an online DeePC framework that incorporates real-time data updates into the Hankel matrix, leveraging the minimum non-zero singular value to selectively integrate informative signals. This approach effectively captures the dynamic nature of the system, ensuring improved control performance. Furthermore, we introduced a numerical SVD technique to mitigate the computational complexity associated with data integration. Simulation results validated the efficacy of the proposed online reduced-order DeePC framework, demonstrating its potential for achieving optimal control in evolving system environments.
Future work will focus on developing criteria to determine the singular value threshold. In addition, we plan to apply the online DeePC framework to complex time-varying systems such as fast charging control of lithium-ion batteries.


\bibliographystyle{ieeetr}
\bibliography{References.bib}
 
\end{document}